# Spectral unmixing of Raman microscopic images of single human cells using Independent Component Analysis

M. Hamed Mozaffari[*a] and Li-Lin Tay[a]

[a]*National Research Council Canada, Metrology Research Centre, Ottawa, ON, Canada.*

[*]*Corresponding author: email:* mhamed.mozaffarimaaref@nrc-cnrc.gc.ca

## Abstract

Application of independent component analysis (ICA) as an unmixing and image clustering technique for high spatial resolution Raman maps is reported. A hyperspectral map of a fixed human cell was collected by a Raman micro spectrometer in a raster pattern on a 0.5-μm grid. Unlike previously used unsupervised machine learning techniques such as principal component analysis, ICA is based on non-Gaussianity and statistical independence of data which is the case for mixture Raman spectra. Hence, ICA is a great candidate for assembling pseudo-colour maps from the spectral hypercube of Raman spectra. Our experimental results revealed that ICA is capable of reconstructing false colour maps of Raman hyperspectral data of human cells, showing the nuclear region constituents as well as subcellular organelle in the cytoplasm and distribution of mitochondria in the perinuclear region. Minimum preprocessing requirements and label-free nature of the ICA method make it a great unmixed method for extraction of endmembers in Raman hyperspectral maps of living cells.

## 1 Introduction

Raman microspectroscopic is becoming more and more popular as a medical imaging modality to study and analysis of living human tissues. The main advantages of Raman imaging over other techniques is the high achievable spatial resolution of Raman microscopy, which is on the same order as visible microscopy [1]. Furthermore, compared to other micro-spectroscopic techniques, Raman imaging does not require any external molecular labels or dyes [2, 3]. However, the differences in Raman spectra from different regions of a human cell are generally quite subtle and make it difficult to only interpret the differences by visualizing Raman spectra and its peaks. Recently, fully-automatic multivariate machine learning (ML) methods have been successfully employed for reconstructing and clustering of Raman images collected from individual human cells [4]. In fact, Raman micro-spectroscopy coupled with ML techniques can be considered as a non-invasive, fast, and automatic imaging technique to utilize for future applications in living cells and in vitro.

Multivariate ML methods use the entire spectrum collected at each x-y plane coordinate to reconstruct a false colour image of the cell under investigation. In other word, the aim of ML methods is to reconstruct an image from a three-dimensional tensor of Raman spectra. Therefore, although univariate methods, including visualizing specific band intensity or intensity ratios, reveal some information on the distribution of biochemical constituents, they cannot rival multivariate techniques in the analysis of Raman hyperspectral data sets. Application of several unsupervised ML methods have been investigated for extraction and clustering constituents of human cell, including agglomerative hierarchical clustering analysis (HCA) [1, 3, 5], k-means clustering analysis (KCA) [6], divisive correlation cluster analysis (DCCA) [4] and fuzzy c-means analysis (FCA) [4, 7]. Similarly, hyperspectral unmixing methods, such as principle component analysis (PCA) [4, 7], vertex component analysis (VCA) [8], and N-Finder [9], have been utilized to extract significant pure Raman spectra (also known as endmembers) from hyperspectral Raman data.

Previous studies revealed that conventional clustering techniques such as KCA perform image reconstruction better than dimensionality reduction and unmixing methods such as PCA or VCA [3, 4]. When we are performing statistical analysis, one assumption is that there is a probability distribution that probabilities of random variables can be drawn from. For many applications, specifically natural phenomena, Gaussian (also called Normal) distribution is a sensible choice. However, there are many situations in which Gaussianity of data not hold [10]. Electroencephalogram (EEG), electrical signals from different parts of scalp, natural images, or human speech are all examples not normally distributed. Raman spectra is another example of signals with no Gaussian distribution [11]. For this reason, methods such as PCA which assumes data as both normal and uncorrelated are weak in extracting components of Raman mixture spectra [12]. Fortunately, unlike PCA, independent component analysis (ICA) is powerful in extracting individual signals from mixtures of signals based on two assumptions for data points, mutually statistically independent and non-Gaussianity [11, 13, 14]. In this study, we have demonstrated that ICA algorithm, as a multivariate method applied to micro-spectroscopic Raman data, can distinguish the nucleus, nucleoli, cytoplasm, and mitochondria of intact fixed cells by their biochemical compositions.

## 2 Independent Component Analysis

Independent component analysis (ICA) is one of the most powerful analytical techniques in blind source separation (BSS) [10, 15]. A famous application of ICA is the "cocktail party problem", where the underlying human speech signals are separated from a sample data consisting of people talking simultaneously in a room [16]. ICA method has been developed to retrieve unknown pure underlying components from a set of linearly mixed signals. Feasibility of ICA method for extraction of pure signals from mixtures is based on two important assumptions [10]. First, the pure components are statistically fully independent rather than just not correlated in PCA. It means that the variation of one pure signal has no influence on the variations of the other pure signals. For the case of Raman spectroscopy, this assumption is fair when pure spectra has no influence on variation of other pure spectra in a mixture spectrum [11]. The second assumption is that ICA consider the distribution of data as non-Gaussian. In general, the main idea of ICA is to perform dimensionality reduction to map data in terms of independent components (ICs).

A Raman spectrometer connected to a confocal microscope with a motorized stage or a scanning mirror acquires Raman microscopic images in a form of three-dimensional tensors $X_{N_x \times N_y \times N_\nu}$ where $N_x \times N_y$ denotes the number of pixels in the x-y plane and $N_\nu$ is the number of Raman shifts in each spectrum. To use matrix algebra algorithms, hyperspectral tensor data can be unfolded into a matrix format $X_{N_x N_y \times N_\nu}$ with $N_x N_y$ samples and $N_\nu$ Raman shifts in each spectrum. Considering that the Raman hyperspectral data consists of spectra generated from linearly combination of

pure Raman spectra and additive noise, the ICA linear model can be written as the following expression:

$$X_{N_x N_y \times N_v} = A_{N_x N_y \times Ic} S_{Ic \times N_v} + E_{N_x N_y \times N_v} \quad (1)$$

where $X$ is the matrix of the observed Raman spectra, $A$ is the mixing or concentration matrix, $S$ is the matrix of pure source Raman spectra that contains ICs, and $E$ is the error matrix. From the main assumptions of ICA, columns of $A_{:,Ic}$ are independent and components in $S$ are mutually statistically independent. On the other hand, significant source spectra have a definite structure (spectra have non-overlapping peaks [11]), and so their intensity does not have a Gaussian distribution [17]. For a noise-free ICA model, the objective is to estimate an unmixing matrix $W$ that, when applied to $X$, produces the estimated matrix $U$. Mathematically, $W$ should be the inverse of $A$, and $U$ should be equal to $S$:

$$U_{Ic \times N_v} = W_{Ic \times N_x N_y} X_{N_x N_y \times N_v} = W_{Ic \times N_x N_y}(A_{N_x N_y \times Ic} S_{Ic \times N_v}) = S_{Ic \times N_v} \quad (2)$$

Then, the mixing matrix A can then be calculated as:

$$A_{N_x N_y \times Ic} = X_{N_x N_y \times N_v} S^T_{N_v \times Ic}(S_{Ic \times N_v} S^T_{N_v \times Ic})^{-1} \quad (3)$$

Like other BBS techniques, the ICA solutions may present some drawbacks and ambiguities. Different from PCA algorithm, since it is not possible to determine the variances of the ICs (because both $S$ and $A$ are unknown), their order and sign cannot be determined [18]. Therefore, the significant independent component can be any of ICs with either positive or negative sign. Moreover, the choice of ICA algorithm and the number of ICs is also another challenge. Lots of algorithms have been proposed to perform ICA calculations such as FastICA, JADE, and MF-ICA [18]. In this work, FastICA algorithm (explain in [13]), which approximates the differential entropy (denoted as negentropy) to estimate $W$, was used. The FastICA is an iterative approach in which non-Gaussianity of data is maximized by updating the difference entropy of $W$ to minimize mutual information.

In order to determine the optimum number of ICs, we utilized two methods proposed in the literature, the sum of the squared residues (SSR) and Durbin-Watson (DW) tests [19]. To calculate SSR test values, corresponding to each maximum number of ICs ($IC_{max} = [1, 10]$), one ICA model is generated. Using the reconstructed matrix $\hat{X}$ and original observed matrix $X$, values of SSR are calculated by the following expression:

$$SSR = \sqrt{\sum_{i=1}^{N_x N_y} \sum_{j=1}^{N_v} (X_{ij} - \hat{X}_{ij})^2} \quad (4)$$

The ICA model corresponding to a minimal SSR is the model with the optimal $IC_{max}$ value. Similarly, the value of DW criterion is defined as:

$$DW = \frac{\sum_{i=2}^{N_v}(s(i) - s(i-1))^2}{\sum_{i=1}^{N_v}(s(i))^2} \tag{5}$$

Where $s(i)$ is equal to the value of $i$th Raman shift in estimated matrix $S$. The value of the DW test method tends to 0 when there is no noise in the signal, and tends towards 2 if the signal contains only noise. Figure 1 shows the values of SSR test on FastICA models applied to our experimental data. Elbow point in the image can be clearly seen on $IC_{max}=5$. Results of the same testing scenario with DW criterion was shown in Figure 2. The optimum number of ICs from the figure is around 5 which has consistency with results of SSR test method.

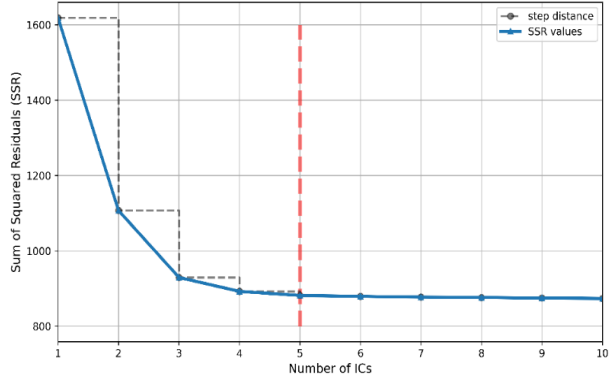

Figure 1. The sum of squared residuals between the original and the reconstructed matrix $X$ for FastICA models with increasing number of ICs.

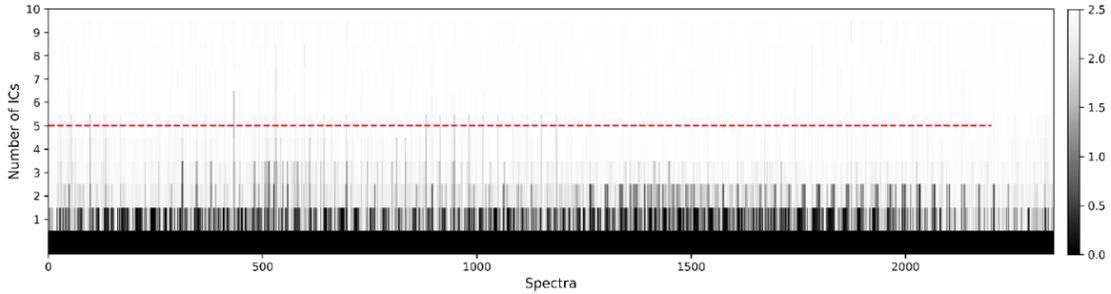

Figure 2. The DW plot: Durbin-Watson values for the residues after subtracting signals calculated using FastICA models with increasing number of ICs.

## 3   Raman data acquisition and preprocessing

It is vital to preprocess data by "whitening" strategy before the application of ICA algorithm [13, 18]. This means that, the observed data are maps linearly to a new space so that its components are uncorrelated and their variances equal unity. In other word, the covariance matrix of mapped observed data equals the identity matrix. Due to the pre-whitening of observed data, the impact of the additive noise on extraction is severely mitigated [13]. Hence, ICA is extremely robust to additive noise in the mixed Raman spectra. Our experimental results indicate that other pre-processing enhancements such as denoising with low-pass filters and baseline correction have minimum impact on the results of ICA. For this reason, preprocessing was kept to a minimum with just whitening and normalization of data.

How the database was acquired?

## 4   Results and discussion

To identify subcellular features of HeLa cell, such as the nucleus, nucleoli, cytoplasm, and mitochondrial regions, ICA was performed in the whole spectral region. The major component in all these three regions is protein. From previous section, the optimum number of clusters for ICA determined to be five. We implemented our ICA model using publicly available SciKit-Learn Python library on a 64-bit Windows PC with 20 CPU cores, 192GB of memory, and NVidia TITAN RTX GPU. The results of ICA in the form of five ICs is displayed in Figure 3. It is noteworthy to mention again that unlike other components analysis technique such as PCA, output set of ICs is not in order of the most significant components as well as ICs sign is not defined. Therefore, negative versions of ICs are also shown, might provide useful information. The image clearly presents different regions in cytoplasm (red in Figure 3c), regions that correlate with a high concentration of mitochondria (yellow in Figure 3h) and the nuclear regions consist of nucleus and nucleoli, respectively (yellow in Figure 3g and d).

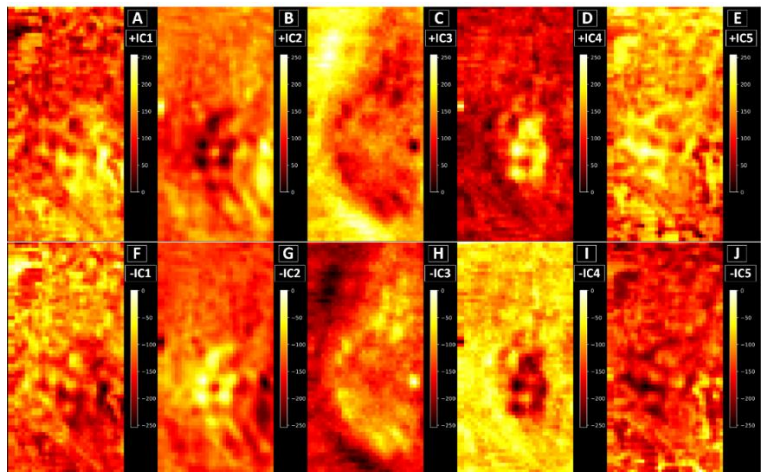

Figure 3. Independent components (ICs) generated by ICA technique from Raman microscopy of HeLa cell. (A-E) Five positive ICs. (F-J) Corresponding negative of the five ICs. Each IC scaled to [0= black, 255=yellow] for the sake of better visualization.

To better distinguishing different nuclear and cytoplasmic regions, a thresholding process was performed on ICs. Figure 4 shows white light image of the cell, an intensity plot of the 2935 cm$^{-1}$ band as a univariate method, the concatenation of ICs after thresholding as well as two ICs related with a different color map for better illustration of cell organelle. Figure 4c shows clearly segmented cytoplasmic regions in green (also light green and blue in +IC2 on Figure 4d), high concentration of mitochondria in red (also yellow, orange, and red in +IC2 on Figure 4d), and nuclear regions in light and dark blue (also yellow, orange, and red in +IC3 on Figure 4e).

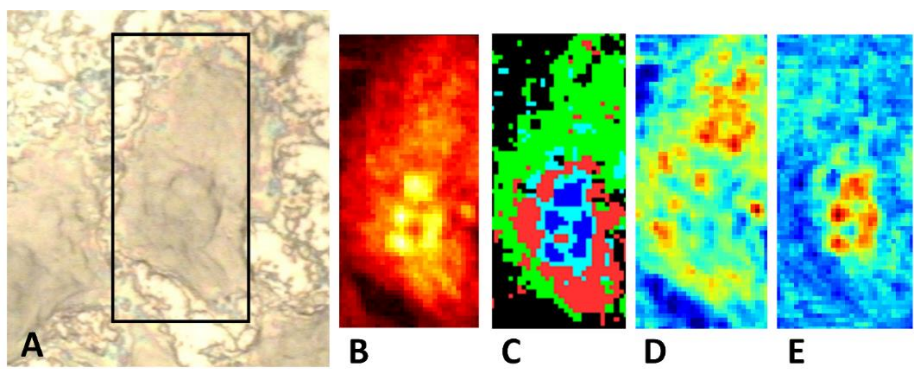

Figure 4. (A) Visual photomicrograph of a HeLa cell. (B) Raman spectral image constructed from only Raman shift 2935 cm$^{-1}$ which has the sharpest Raman peak. (C) Segmentation map of the same cell using based on results of ICA. (D & E) +IC2 and +IC3 in different color map.

As the ICs are estimated from Raman spectra based on wavenumbers and their intensity variations, the average cluster spectra corresponding to each distinguished region contain valuable information of the underlying biochemical differences. Several hundred of spectra contribute to each mean cluster spectra, resulting in very good signal to noise ratio for the mean cluster spectra, allowing interpretation of these spectra and the spectral differences corresponding to different regions in the cell.

Mean spectra representing four clusters are shown in Figure 5. ICA was performed in the spectral range between 1050 and 1800 cm$^{-1}$ region which exhibits the most predominant protein peaks. Spectral information in this range contained sufficient spectral information to give the best clustering result. Furthermore, ICA was performed on the C-H stretching region between 2800 and 3400 cm$^{-1}$. Trace 1 and 2 represent the nuclear region consist of nucleoli and nucleus, respectively (dark and light blue in Figure 4c), trace 3 the perinuclear regions contain regions with high concentration of mitochondria (red in Figure 4c), and trace 4 the region containing cytoplasm (green in Figure 4c). Main regions related to the contribution of proteins can be seen in all traces with distinct bands at 1655 cm$^{-1}$ (amide I vibration) and the extended amid III region between 1250 and 1350 cm$^{-1}$ (Peptide backbone and coupled C-H, N-H deformation modes). Furthermore, distinct peaks, which can be related to deformations of saturated and unsaturated fatty acid side chains, appear between 1200 and 1350 cm$^{-1}$.

In Raman bands between 2850 and 2900 cm$^{-1}$, spectral differences of the C-H-stretching regions (CH$_3$, CH$_2$, and CH), more related to the cytoplasm and the mitochondrial regions are distinguishable. In fact, the rise of intensities between 2850 and 2935 cm$^{-1}$ is given by relatively long alkane chains of lipids. The spectral difference is better distinguishable for cytoplasm (trace 4) and perinuclear regions (trace 3) compared with the nucleus (trace 1 and 2). Antisymmetric methyl and methylene deformations, peptide side chains, and phospholipids contribute to all traces by a sharp peak in region between 1425 and 1475 cm$^{-1}$. A rising shoulder at region between 1050 and 1100 cm$^{-1}$ is related to symmetric stretching mode of phosphate esters, DNA, RNA, and phospholipids. Fingerprints of other membranous cell organelles in perinuclear area could be weakly seen in trace 3, including endoplasmic reticulum with ribosomes, the Golgi apparatus, mitochondria, lysosomes, intracellular vesicles, cholesterol, phospholipids, and fatty acids. Due to the significant contribution of protein bands in all Raman spectra, subtracting cytoplasm and nucleus Raman spectra might provide a better visualization of other constituents. This difference spectrum is shown in trace 0 of Figure 5. For instance, negative differences might be associated with the contribution DNA.

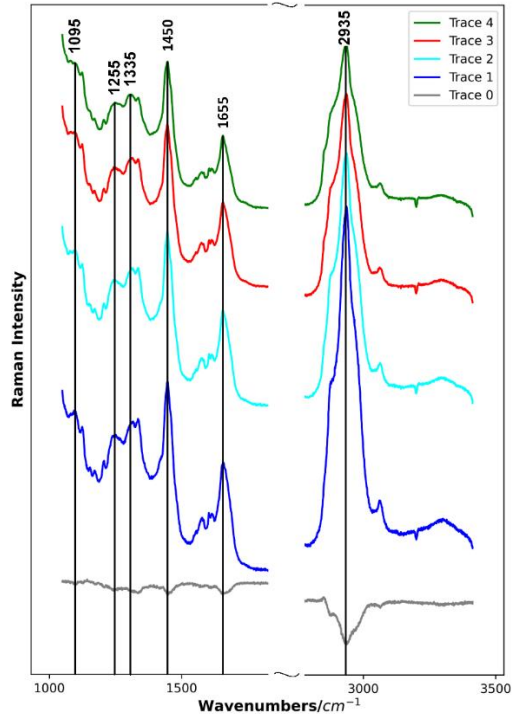

Figure 5. Average spectra from ICA representing nucleoli (trace 1), nucleus (trace 2), perinuclear area with high concentration of mitochondrial (trace 3), cytoplasm (trace 4), and difference spectrum between perinuclear areas and nucleus (trace 0).

## 5   Conclusion

Previously, it has been demonstrated that Raman spectrum of a biological sample might be a linear combination of several pure Raman spectra when they are mutually statistically independent [11, 17]. Also, statistical distribution of Raman spectra are less likely to be Gaussian. These two characteristics have a huge correlation with the two fundamental assumptions for independent component analysis (ICA) technique. Therefore, ICA can be a great unmixing alternative to apply to hyperspectral Raman map to extract source Raman spectra (i.e. endmembers). We have demonstrated that Raman micro-spectroscopy, coupled to a multivariate unsupervised machine learning technique such as ICA, can distinguish the nucleus, nucleolus, cytoplasm, and mitochondria of human call by their biochemical compositions. This novel idea provides a new tool for non-invasive, in vitro studies of cell biological aspects without using any external labels.